\documentstyle[twoside,fleqn,espcrc2,epsf]{article}
\newcommand{\bra}[1]{\left< #1 \right|}

\newcommand{\bket}[1]{\langle#1\rangle}
\newcommand{\ket}[1]{\left| #1 \right>}
\newcommand{\dla}{\stackrel{\leftarrow}{D}}

\newcommand{\eq}{\begin{equation}}
\newcommand{\en}{\end{equation}}
\newcommand{\eqa}{\begin{eqnarray}}
\newcommand{\ena}{\end{eqnarray}}

\newcommand{\AmS}{{\protect\the\textfont2
  A\kern-.1667em\lower.5ex\hbox{M}\kern-.125emS}}

\title{Aspects of determining $f_{B_s}$: scaling and power-law divergences} \author{Presented by
  S.~Collins, UKQCD Collaboration\address{Dept. of Physics an d Astronomy,
    Glasgow University, Glasgow, G12 8QQ, Scotland} \thanks{In collaboration
    with C.~T.~H.~Davies and J.~Hein, Glasgow Univ.; J.~Shigemitsu, Ohio-state
  Univ.; C.~Morningstar, San Diego; A.~Ali Khan, Tsukuba.}}

\begin{document}

\begin{abstract}
  We present {\it preliminary} results for the decay constant of the
  $B_s$ meson, $f_{B_s}$, at three values of $\beta=5.7$, $6.0$ and
  $6.2$ using NRQCD and clover fermions for the heavy and light quarks
  respectively.  As a consistency check the decay constant has also
  been extracted from the axial-vector current at finite momentum. In
  addition, we discuss the cancellation of $O(\alpha/(aM_0))$ terms and
  the remaining uncertainty in $f_{B_s}$ from higher order
  divergences.
\end{abstract}

\maketitle

\section{Simulation details}
For the heavy quark we use an NRQCD action consistent to $O(1/M_0^2)$:
\begin{equation}
S = Q^\dagger(\Delta_t+H_0+\delta H)Q
\end{equation}
where,
\begin{eqnarray}
H_0 & = & -\frac{\Delta^{(2)}}{2M_0}\\
\delta H & = & -c_1\frac{{\mathbf \sigma}\cdot{\mathbf B}}{2M_0} +c_2\frac{({\mathbf \Delta\cdot E} - {\mathbf E\cdot\Delta})}{8M_0^2} \nonumber\\
         &   & -c_3 \frac{{\mathbf \sigma\cdot}({\mathbf \Delta\times E} - {\mathbf
           E\times\Delta})}{(8M_0^2)}-c_4\frac{({\mathbf \Delta}^{(2)})^2}{8M_0^3} \nonumber\\
         &   &  + c_5\frac{a^2{\mathbf
             \Delta}^{(4)}}{24M_0}-c_6\frac{a({\mathbf \Delta}^{(2)})^2}{16nM_0^2}.
\end{eqnarray}
The $O(1/M_0^3)$ correction to the kinetic energy~(expected to be the largest
contribution from this order) and the first two discretisation
corrections are also included.  We implement tadpole improvement throughout,
using the plaquette definition of $u_0$, and set $c_i=1$.  

For the light quark we use the clover action with the tadpole improved value
of $c_{SW}$ at $\beta=5.7$ and $6.0$ and the non-perturbatively determined
value at $\beta=6.2$; $c_{SW}=1.61$ as determined by the Alpha
collaboration~\cite{nonpert}, compared to $c_{SW}=1.48$ using
tadpole-improvement. The configurations at $\beta=5.7$ and the configurations
and light quark propagators at $\beta=6.2$ were generously provided by the
UKQCD collaboration. The light quark mass is fixed to the strange quark mass
using the $K$ meson mass, with the uncertainty in this determination estimated
by fixing $m_q$ using the $\phi$. Further simulation details for the 3
ensembles are given in table~\ref{simdet}. 

\begin{table}
\begin{tabular}{|c|c|c|c|c|}\hline
$\beta$ & V & N & $a^{-1}(m_\rho)$~(GeV) & $aM_0^b$\\\hline
5.7 & $12^3\times 24$ & 278 & 1.14(3) & $\sim 4.2$\\
6.0 & $16^3\times 48$ & $102\time 2$ & 1.92(7) & $\sim 2.2$\\
6.2 & $24^3\times 48$ & 144 & 2.63(9) & $\sim 1.6$\\\hline
\end{tabular}
\caption{The simulation details. The errors on $a^{-1}$ include statistical errors
  and those due to the chiral extrapolation of $m_\rho$. N denotes the number of configurations.  }\label{simdet}
\end{table}
The pseudoscalar decay constant is defined as
\begin{equation}
\bra{0}A_{\mu}\ket{PS}_{QCD} = p_{\mu} f_{PS}
\end{equation}
in Euclidean space. On the lattice matching factors $C_i$ relate the lattice
operators to the current in full QCD. For the zeroth component of the current
to $O(\alpha/M)$:
\begin{equation}
\bket{A_0}_{QCD} = \sum_j C_j(\alpha,aM_0)\bket{J_L^{i}}
\label{current}
\end{equation}
where,
\begin{eqnarray}
O(1): J_L^{(0)} & = & \bar{q}\gamma_5\gamma_0 Q\\
O(\frac{1}{M}): J_L^{(1)} & = & -\frac{1}{2M_0}\bar{q}\gamma_5\gamma_0 ({\mathbf \gamma\cdot
  D}) Q\\
O(\frac{\alpha}{M}): J_L^{(2)} & = & \frac{1}{2M_0}\bar{q}({\mathbf \gamma\cdot
  \dla})\gamma_5\gamma_0  Q.
\end{eqnarray}
The $O(a\alpha)$ discretisation error in the current is removed by
defining~\cite{pert},
\begin{eqnarray}
J_L^{0,imp} & = & J^{(0)}_L+C_A J^{(disc)}_L\\
J_L^{(disc)}& = & a\bar{q}({\mathbf \gamma\cdot   \dla})\gamma_5\gamma_0  Q
\end{eqnarray}
The $C_i$s have been calculated to 1-loop~\cite{pert}, however, the
$q^*$~(which also depends on $aM_0$) at which the strong coupling, $\alpha$,
is computed is not yet known. Thus, we average the results obtained using
$aq^*=1.0$ and $\pi$. Note that in the static limit $aq^*\sim2$~\cite{static}
for Wilson light fermions.

The results at $\beta=5.7$ and $6.0$ have appeared previously
in~\cite{jhein} and~\cite{alikhan} respectively. Further details of our
methods and analysis can be found in these references.

\section{Scaling of $f_{B_s}$}
\begin{figure}
  \vskip -0.5cm \epsfxsize=6cm\epsfbox{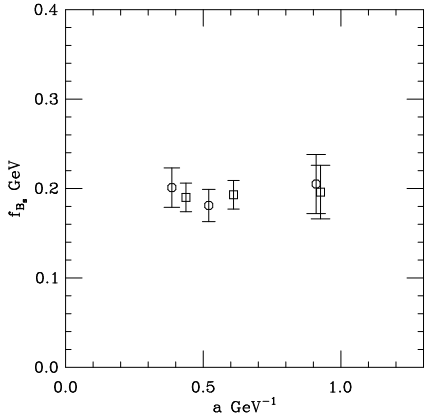} \vskip-1cm
\caption{Preliminary results for $f_{B_s}$ as a function of the lattice
  spacing. Our results~(circles) are compared to those from
  JLQCD~(squares)~\protect\cite{jlqcd}. All errors include statistical and
  systematic uncertainties added in quadrature.}
\label{scale}
\vskip -0.5truecm
\end{figure}

Our values for $f_{B_s}$, calculated to $O(\alpha/M)$, are presented in
figure~\ref{scale} as a function of $a^{-1}$. The results are consistent with
scaling, and $f_{B_s}\sim 200$~MeV. Nice agreement is found with the results
of the JLQCD collaboration~\cite{jlqcd}, also shown in the figure. This group
uses the string tension to set $a$, and an $O(1/M_0)$ NRQCD action. However,
this is unlikely to affect the comparison significantly.

Table~\ref{errors} details how the errors are estimated. Note that $a^{-1}\sim
1{-}2.6$~GeV is the range in which the NRQCD action and clover light fermions can
be applied to the $B$ meson. For coarser values of $a$, the discretisation
errors from the light quark action~(and gauge action) increase rapidly, as do
the $O(\alpha^2)$ perturbative errors.  Conversely, if $a$ becomes too fine,
the discretisation errors are under control but the $O(\alpha/(aM)^2)$
perturbative errors increase dramatically as $aM_0^b$ drops below 1. Overall,
the perturbative errors are the main source of uncertainty, although at
$\beta=5.7$ the discretisation errors are of equal magnitude.

\begin{table}
\begin{tabular}{|c|c|c|c|}\hline
Source & $\beta=5.7$ & $6.0$ & $6.2$\\\hline
statistical &  $3$ & $3$ & $2$\\
disc. $O((a\Lambda_{QCD})^2)$ & {\bf 13}    & 4   & 2  \\
pert. $O(\alpha^2,\alpha/(aM)^2)$ & {\bf 13}    &  {\bf 8}  & {\bf 9}  \\
NRQCD $O(1/M^2)$  & 1    &  1  &  1 \\
$\kappa_s$  &  +4   & +4   & +4  \\
$a^{-1}(m_\rho)$ &  3   &  4  &  3 \\\hline
Total &  {\bf 19} & {\bf 11} & {\bf 11}\\\hline
\end{tabular}
\caption{Estimates of the statistical and main systematic errors, in percent, in our values for $f_{B_s}$}\label{errors}
\vskip -0.7truecm
\end{table}
\section{$f_{B_s}$ extracted at finite momentum}
In order to investigate momentum dependent discretisation errors in
the decay constant we computed the ratio
\begin{equation}
\bket{J^{(0)}_L}_{\vec{p}}/\bket{J^{(0)}_L}_{\vec{0}} = \sqrt{E(p)/M}\label{ratmom}
\end{equation}
where $\bket{J^{(i)}_L}_{\vec{p}}=f^{i}\sqrt{E(\vec{p})}$~(without
renormalisation).  The RHS of equation~\ref{ratmom} is a slowly
varying function of $|\vec{p}|$, which is close to $1$ for the range
of momenta we studied~(up to $1.5$~GeV at $\beta=6.2$). Our results,
presented in table~\ref{momenta}, are in agreement with this
expectation at $\beta=6.0$ and $6.2$, i.e. the momentum dependent
discretisation errors in $f_{B_s}$ are not significant. However, at
$\beta=5.7$ a $10{-}20\%$ deviation from $1$, is seen as $|\vec{p}|$
increases, although this is within the magnitude expected for
$O((ap)^2)$ discretisation errors.
\begin{table}
\begin{tabular}{|c|c|c|c|c|}\hline
\multicolumn{5}{|c|}{$(f^{0}\sqrt{E(\vec{p})})/(f^{0}\sqrt{M})$}\\\hline
$\beta$ & $n^2= 1$ & 2 & 3 & 4 \\\hline
 {\small 5.7} &
{\small 0.94(1)} & {\small 0.88(1)} & {\small 0.84(1)} & {\small 0.82(2)}\\
 {\small 6.0} & {\small 1.01(1)} & {\small 1.01(3)} &- &- \\
 {\small 6.2} & {\small 1.00(2)} & {\small 1.02(3)} & {\small 1.03(4)} & {\small 1.07(6)} \\\hline
\end{tabular}
\caption{The decay constant extracted at
  finite momentum for $aM_0$ close to $aM_0^b$. $|\vec{p}|=2n\pi/(aL)$, where
  $n=0,1,\protect\sqrt{2}\ldots$ and $L$ is the spatial extent of the lattice.
}\label{momenta}
\vskip -0.5truecm
\end{table}
\section{Power-law divergences}
Matrix elements in NRQCD beyond zeroth order diverge as $aM_0\rightarrow 0$.
In the case of $f_B$ there are unphysical, ultra-violet, contributions to
$\bket{J_L^{(i)}},i>0$, which are cancelled order by order in perturbation
theory by terms appearing in the perturbative coefficients. In general,
simulations are performed using $aM_0{>}1$ and so the unphysical contributions
are not expect to be large and their cancellation should be under control.

Considering equation~\ref{current} in more detail~(see~\cite{pert} for definitions), 
\begin{eqnarray}
\bket{A_0}_{QCD} & = & (1+\alpha\rho_0)\bket{J_L^{(0)}}+ \bket{J_L^{(1)}} \nonumber\\
& & \hspace{-2.cm}+\alpha\rho_1\bket{J_L^{(1)}} 
+\alpha\rho_2\bket{J_L^{(2)}}\nonumber\\
& &\hspace{-2.cm} + \alpha\rho_{disc}\bket{J_L^{(disc)}}\label{detail}
\end{eqnarray}
The explicit contributions to $\rho_0$ are
\begin{equation}
\rho_0 = [B_0-\frac{1}{2}(C_q+C_Q)-\zeta_{00}-\zeta_{10}].
\end{equation}
The lowest order divergent contribution to the current is $O(\alpha/(aM))$,
which appears through the tree-level term $\bket{J_L^{(1)}}$ and is {\bf
  cancelled} by the mixing term $\alpha\zeta_{10}\bket{J_L^{(0)}}$, where
$\zeta_{10}$ is the renormalisation due to the mixing between $J_L^{(0)}$ and
$J_L^{(1)}$.  The remaining divergent contributions,
$O(\alpha/(aM)^2,\alpha^2/(aM))$ etc, appearing in equation~\ref{detail}, are
cancelled at higher orders; the uncertainty in $f_{B_s}$ due to these
remaining terms is well within our estimates of the systematic uncertainties
in table~\ref{errors}.

\begin{figure}
\vskip -0.5cm
\epsfxsize=6cm\epsfbox{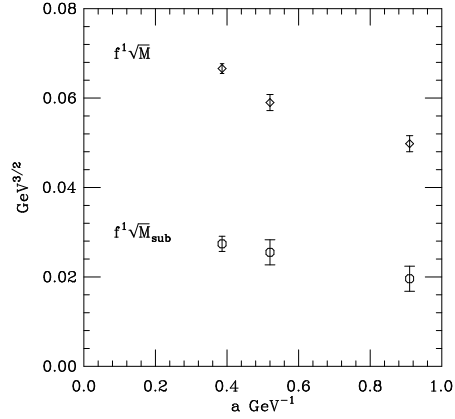}
\vskip-1cm
\caption{The contribution to the decay constant from the
  tree-level current correction before and after the subraction of the
  $\alpha\zeta_{10}\bket{J_L^{(0)}}$, $aq^*=2.0$.}
\label{wave}
\vskip -0.5truecm
\end{figure}

In fact a large part of $\bket{J_L^{(1)}}$ is unphysical,
$\alpha\zeta_{10}\bket{J_L^{(0)}}/\bket{J_L^{(1)}}\sim 0.5-0.6$ for all
$\beta$s and $aM_0$ for $aq^*=2.0$. However, once the $O(\alpha/(aM))$
contribution to $\bket{J_L^{(1)}}$ is cancelled, the scaling behaviour of this
term improves, as shown in figure~\ref{wave}, suggesting the remainder is
physical.  However, $\bket{J_L^{(1)}}-\alpha\zeta_{10}\bket{J_L^{(0)}}\sim 0.025\mbox{GeV}^{3/2}$ is
small,~(cf $f^0\sqrt{M}\sim 0.500\mbox{GeV}^{3/2}$), and of the order of the higher
order perturbative and NRQCD corrections. Hence, we cannot determine the size
of the physical part of $\bket{J_L^{(1)}}$ reliably. We emphasise this does
not lead to a significant uncertainty in $f_{B_s}$, nor in the slope of the
decay constant with $1/M$, which are dominated by $\bket{J_L^{(0)}}$ and the
corresponding perturbative coefficient~(without the $\zeta_{10}$ term).

\section*{ACKNOWLEDGEMENTS}
\noindent
S.~Collins has been supported by a Royal Society of Edinburgh fellowship.

\end{document}